\documentstyle[preprint,aps]{revtex}
\begin{document}
\draft
\title{Pseudoclassical model for Weyl particle in 10 dimensions}
\author{D. M. Gitman, and A. E. Gon\c {c}alves.}
\address{Instituto de F\'{\i}sica, Universidade de S\~ao Paulo\\
P.O. Box 66318, 05389-970 S\~ao Paulo, SP, Brazil}
\date{\today}
\maketitle
\begin{abstract}
A pseudoclassical model to describe Weyl particle in 10 dimensions
is proposed. In course of quantization both the massless Dirac equation and
the Weyl condition are reproduced automatically.
The construction can be relevant to Ramond-Neveu-Schwarz strings
where the Weyl reduction in the Ramond sector has to be made by hands.
\end{abstract}
\pacs{PACS number(s): 11.10.Ef, 036.65.Pm}

Classical and pseudoclassical models of relativistic particles and
their quantization attract attention already for a long time. One of the main
reason to study them is related to the string theory, because  point-like
particles can be treated as prototypes of strings or some modes in string
theory. Recently \cite{GGT}, a pseudoclassical model to describe Weyl
particle in 4 dimensions was proposed. The limit $m\rightarrow 0$ of the
standard action  of a spinning particle \cite{b2} in 4 dimensions was
modified essentially to get the minimal theory (Weyl theory) of a massless
spinning particle. Thus, both Dirac equation and Weyl
condition are reproduced in course of a quantization. It turns out to
be possible to adopt the model to 10 dimensions. That is important in
connection with superstring theory problems in such dimensions
\cite{GS}, where, for example, the minimal quantum theory of
Ramond-Neveu-Schwarz string does not appear automatically, and the
corresponding GSO reduction (in particular Weyl reduction of the Ramond
sector) has to be made by hands \cite{b4}.

The action of the Weyl particle in 10 dimensions we are proposing has the form
\begin{eqnarray}\label{AWp}
S=\int_0^1\left[
-\frac{z^2}{2e}-\imath\psi_\mu\dot{\psi}^\mu
\right]d\tau\;,\;\;
z^\mu =\dot{x}^\mu-\imath\psi^\mu\chi-
\frac{1}{8!}\epsilon^{\mu\nu\rho_2\ldots\rho_9}
b_\nu\psi_{\rho_2}\ldots\psi_{\rho_9} +\frac{\alpha}
{2^8}b^\mu\;,
\end{eqnarray}
where $x^\mu ,\;e$ are even and $\psi^\mu ,\;\chi $ are odd variables
dependent on a parameter $\tau\;\in\;[0,1],\;\mu = \overline{0,\,9},\;
\eta_{\mu\nu}={\rm diag}(1,-1,\ldots , -1)$ is Minkowski tensor in 10
dimensions, the variables $b^\mu$ form an
even $10$-vector, and $\alpha$ is an even constant.
There are three types of gauge transformations under which the action
(\ref{AWp}) is invariant: reparametrizations
\begin{equation}\label{repar}
\delta x^{\mu} =  {\dot{x}}^{\mu} \xi\;,\;\;
\delta e = \frac{d}{d\tau}\left(e \xi\right)\;,\;\;
\delta b^\mu = \frac{d}{d\tau}\left(b^\mu\xi\right)\;,\;\;
\delta \psi^\mu = \dot{\psi}^\mu\xi\;,\;\;
\delta\chi = \frac{d}{d\tau}\left(\chi\xi\right)\;,
\end{equation}
with an even parameter $\xi(\tau)$; and two kinds of supertransformations:
first ones
\begin{eqnarray}\label{supert}
&&\delta x^{\mu}  =  \imath\psi^{\mu}\epsilon\;,\;\;
\delta e =\imath\chi\epsilon\;,\;\;
\delta b^\mu  = 0\;,\;\;
\delta \psi^{\mu}  =
\frac{1}{2e}z^\mu\epsilon\;,\;\;
\delta \chi= \dot{\epsilon}\;,
\end{eqnarray}
with an odd parameter $\epsilon (\tau)$; and second ones
\begin{eqnarray}\label{gauge}
&&\delta x^\mu = \left(\frac{1}{8!}\epsilon^{\mu\nu\rho_2\ldots\rho_9}
b_\nu\psi_{\rho_2}\ldots\psi_{\rho_9}-\frac{\alpha}{2^8}b^\mu\right)
\kappa\;,\;\;
\delta\psi^\mu =
\frac{\imath}{8!e}\epsilon^{\mu\nu\rho_2\ldots\rho_9}b_\nu
z_{\rho_2}\psi_{\rho_3}\ldots\psi_{\rho_9}\kappa\;,\nonumber\\
&&\delta b^\mu = \frac{d}{d\tau}(b^\mu\kappa)-\left(\frac{6}{\alpha}\right)
\frac{2^8}{8!}\epsilon^{\mu\nu\rho_2\ldots\rho_9}\dot{\psi}_{\rho_2}
\psi_{\rho_2}\ldots\psi_{\rho_9}\kappa\;,\;\;
\delta \chi = 0\;,\;\;\delta e = 0\;,
\end{eqnarray}
with an even parameter $\kappa(\tau)$.

Introducing the canonical momenta
\begin{eqnarray*}
&&\pi_{\mu}=\frac{\partial L}{\partial \dot{x}^{\mu}} =
-\frac{1}{e}z_\mu\;,\;
P_{e}= \frac{\partial L}{\partial \dot{e}} = 0\;,\;
P_{\chi} = \frac{\partial_{r}L}{\partial \dot{\chi}}= 0\;,\\
&&P_\mu = \frac{\partial_{r}L}{\partial \dot{\psi}^\mu} =
-\imath\psi_\mu\;,\;
P_{b_\mu}=\frac{\partial L}{\partial \dot{b^\mu}} =0\;,
\end{eqnarray*}
we discover primary constraints $\Phi^{(1)}=0\; (
\Phi^{(1)}_{1} = P_e\; ,\;\;
\Phi^{(1)}_{2} = P_{\chi}\;,\;
\Phi^{(1)}_{3 \mu} = P_\mu+\imath\psi_\mu\;,\;
\Phi^{(1)}_{4 \mu} = P_{b_\mu})$.
Then the total Hamiltonian $H^{(1)}$ constructed according
to the standard procedure \cite{Dirac,DI}, has the form
$H^{(1)} = H+\lambda_{A}\Phi_{A}^{(1)}\;,$ where
\begin{equation}\label{H}
H = -\frac{e}{2}\pi^2+\imath\pi_\mu\psi^\mu\chi-\left(
\frac{1}{8!}\epsilon_{\nu\mu\rho_2\ldots\rho_9}\pi^\mu
\psi^{\rho_2}\psi^{\rho_9}+
\frac{\alpha}{2^8}\pi_\nu\right) b^\nu\;.
\end{equation}
Using it, one gets secondary constraints $\Phi^{(2)}=0$,
\begin{equation}\label{SC}
\Phi^{(2)}_1 = \pi^2\;,\;\;
\Phi^{(2)}_2 = \pi_\mu\psi^\mu\;,\;\;
\Phi^{(2)}_{3\mu} =
\frac{1}{8!}\epsilon_{\mu\nu\rho_2\ldots\rho_9}\pi^\nu\psi^{\rho_2}
\ldots\psi^{\rho_9}+\frac{\alpha}{2^8}\pi_\mu\;.
\end{equation}
One can go over from the initial set of
constraints ($\Phi^{(1)}\;,\Phi^{(2)}$) to the equivalent one
($\Phi^{(1)}\;,\widetilde{\Phi}^{(2)}$), where
$\widetilde{\Phi}^{(2)} = \Phi^{(2)}\left|_{\psi\rightarrow
\widetilde{\psi} = \psi+\frac{\imath}{2}\Phi_3^{(1)}\;.}
\right.$ The new set of constraints can be explicitly divided in
a set of first-class constraints, which is $(\Phi^{(1)}_{1,2},
\Phi^{(1)}_4, \widetilde{\Phi}^{(2)})$ and in a set of second-class
constraints $\Phi^{(1)}_3$.

Consider the Dirac quantization, where the second-class
constraints $\Phi^{(1)}_3$ defines Dirac brackets and therefore the
commutation relations, whereas, the first-class constraints, being
applied to the state vectors, define physical states \cite{Dirac}.
For essential operators and nonzeroth commutators we get:
\begin{equation}\label{RCDQ}
\left[\hat{x}^\mu,\hat{\pi}_\nu\right]_{-}=\imath
\left\{x^\mu,\pi_\nu\right\}_{D(\Phi^{(1)}_3)}=
\imath\delta^\mu_\nu\;,\;\;
\left[\hat{\psi}^\mu,\hat{\psi}^\nu\right]_{+}=
\imath\left\{\psi^\mu,\psi^\nu\right\}_{D(\Phi^{(1)}_3)}=
-\frac{1}{2}\eta^{\mu \nu}\;.
\end{equation}
It is possible to construct a realization of the commutation relation
(\ref{RCDQ}) in a Hilbert space ${\cal R}$ whose elements $\Psi$ are
32-components columns. Taking into account trivial
first-class constraints $\Phi^{(1)}_{1,2,4}$, we can select $\Psi$ dependent
only on $x$. Then
\begin{equation}\label{REAL}
\hat{x}^\mu = x^\mu {\bf I}\;,\;\;\hat{\pi}_\mu = -\imath\partial_\mu
{\bf I}\;,\;\;\hat{\psi}^\mu = \frac{\imath}{2}\gamma^\mu\;,
\end{equation}
where ${\bf I}$ is $32\times 32$ unit matrix and $\gamma^\mu$ are the
$\gamma$-matrices in 10 dimensions,
$\left[\gamma^\mu\;,\;\gamma^\nu\right]_+=2\eta^{\mu\nu}$.
In the realization (\ref{REAL}) the operators $\hat{\Phi}^{(2)}$, which
corresponds to the first-class constraints (\ref{SC}),
have the following form
\[
\hat{\Phi}^{(2)}_1 =-\partial_\mu\partial^\mu\;,\;\;
\hat{\Phi}^{(2)}_2 =\frac{1}{2}\partial_\mu\gamma^\mu\;,\;\;
\hat{\Phi}^{(2)}_{3\mu} =\frac{\imath}{2^8}
\left[\gamma_\mu\gamma^{11}\partial_\nu\gamma^\nu + \partial_\mu
\left(\gamma^{11} -\alpha\right)\right]\;.
\]
One can see that conditions $\hat{\Phi}^{(2)}\Psi (x)=0$,
are reduced to the following set of independent equations
\[
\partial_\mu\gamma^\mu\Psi (x)=0\;,\;\;
\partial_\mu\left(\gamma^{11}-\alpha\right)\Psi (x)=0\;.
\]
The first one is the Dirac equation for massless particles in 10
dimensions and the second one at $\alpha =\mp 1$ is equivalent to Weyl or
anti-Weyl conditions $\left(1\pm\gamma^{11}\right)\Psi (x)=0$
if we consider only normalized functions $\Psi (x)$. Thus, we get
automatically projections with positive and negative chirality in course
of quantization.

The canonical quantization can be made similar to 4 dimensional case
\cite{GGT} and leads to same quantum mechanics.
\acknowledgments
The authors thank N. Berkovits for helpful discussions and Brazilian
foundations CNPq and CAPES for support.

\end{document}